\documentclass[aps,prb,twocolumn,superscriptaddress,showpacs]{revtex4}

\usepackage{amsmath}

\usepackage{graphicx}

\begin{document}

\title{Condon Domain Phase Diagram for Silver}

\author{R.~B.~G.~Kramer}
\affiliation{LNCMI, CNRS, BP
166, 38042 Grenoble Cedex 9, France}
\affiliation{Max-Planck-Institut f\"{u}r Festk\"{o}rperforschung,
Heisenbergstra{\ss}e 1, 70569 Stuttgart, Germany}
\affiliation{Institut N\'eel, CNRS, Universit\'e Joseph Fourier, BP 166, 38042 Grenoble Cedex 9, France}

\author{V.~S.~Egorov}
\affiliation{LNCMI, CNRS, BP
166, 38042 Grenoble Cedex 9, France}
\affiliation{Max-Planck-Institut f\"{u}r Festk\"{o}rperforschung,
Heisenbergstra{\ss}e 1, 70569 Stuttgart, Germany}
\affiliation{Russian Research Center "Kurchatov Institute", 123182
Moscow, Russia}

\author{V.~A.~Gasparov}
\affiliation{Institute of Solid State Physics, Russian Academy of
Sciences, 142432 Chernogolovka, Russia}

\author{A.~G.~M.~Jansen}
\affiliation{Service de Physique Statistique, Magn\'{e}tisme, et
Supraconductivit\'{e}, INAC, CEA, 38054 Grenoble Cedex 9, France}

\author{W. Joss}
\affiliation{LNCMI, CNRS, BP
166, 38042 Grenoble Cedex 9, France}
\affiliation{Max-Planck-Institut f\"{u}r Festk\"{o}rperforschung,
Heisenbergstra{\ss}e 1, 70569 Stuttgart, Germany}
\affiliation{Universit\'e Joseph Fourier, BP 53, 38041 Grenoble
Cedex 9, France}

\begin{abstract}
We present the Condon domain phase diagram for a silver single
crystal measured in magnetic fields up to 28~T and temperatures down
to 1.3~K. A standard ac method with a pickup coil system is used at
low frequency for the measurements of the de Haas-van Alphen effect
(dHvA). The transition point from the state of homogeneous
magnetization to the inhomogeneous Condon domain state (CDS) is
found as the point where a small irreversibility in the dHvA
magnetization arises, as manifested by an extremely nonlinear
response in the pickup voltage showing threshold character. The
third harmonic content in the ac response is used to determine with
high precision the CDS phase boundary. The experimentally determined
Condon domain phase diagram is in good agreement with the
theoretical prediction calculated by the standard Lifshitz-Kosevich
(LK) formula.
\end{abstract}
\date{\today}
\maketitle

\section{Introduction}

The formation of dia- and paramagnetic domains has been predicted by
Condon~\cite{Condon1966} to occur in non-magnetic pure metals by
considering the collective interaction between the electrons on
Landau-quantized energy levels in the de Haas-van Alphen (dHvA)
effect. These domains corresponding to an inhomogeneous
magnetization are commonly called Condon domains. The domain
formation results from a self-consistent treatment of the
oscillating dHvA magnetization $M$ due to the orbital quantization
of the electronic system in the total magnetic induction $B =\mu_0(
H + M)$, where $M=M(B)$ depends on the total induction $B$ in an
applied magnetic field $H$. Following the Pippard-Shoenberg concept
of magnetic interaction~\cite{Pippard1963,Shoenberg1984} where the
electrons experience the influence of the magnetic field induced
magnetization of all neighboring electrons, a thermodynamic
instability arises when the dHvA amplitude becomes large enough,
i.e., the differential susceptibility
\begin{equation}
\chi =\mu_0\frac{\partial M}{\partial B}>1. \label{equation1}
\end{equation}
This condition corresponds to the situation where the amplitude of
the magnetization amplitude becomes comparable to the magnetic field
period of the dHvA effect. For sufficiently strong magnetization
amplitudes, this instability condition, rewritten like
$\mu_0\partial H/\partial B=1 - \chi < 0$, occurs in a certain field
interval within the paramagnetic part of each dHvA cycle. In these
field intervals, where $\mu_0\partial H/\partial B<0$, the induction
as function of the applied field $B(H)$ is multi-valued, like the
van der Waals isotherm for a real gas. The system avoids this instability in the same way as the real gas. For an infinite long
rod-like sample (demagnetization factor $n=0$), the induction $B$
undergoes a discontinuous transition between the two stable states
with the induction $B_1$ and $B_2$ at a certain applied field $H$, like the liquid-gas specific volumes change discontinuously
at the equilibrium vapor pressure. Both stable states $B_1$ and
$B_2$ correspond to the same free energy and the inductions in the
instability interval ($B_1,B_2$) are never realized.
\begin{figure}[tb]
\begin{center}
       \includegraphics*[width=0.95\linewidth, bb=16 8 114 110]{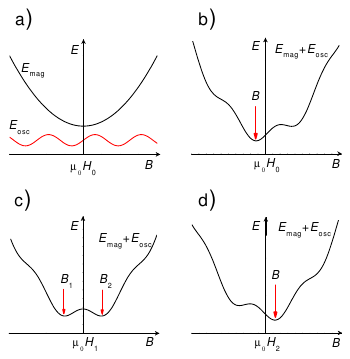}
       \caption{(Color online) (a) Schematic representation of the magnetization energy $E_{mag}$ and the dHvA-energy
$E_{osc}$ as function of $B$ for $n=0$. (b)-(d) Sum of these energies for three applied magnetic fields $H_0 < H_1 < H_2$. If the curvature of the parabola is smaller
than the curvature of the oscillating energy two minima coexist at $H_1$. This leads to discontinuous jump of the induction when sweeping the magnetic field through $H_1$.
       \label{Fig_Energyargument}}
        \end{center}
\end{figure}
Figure~\ref{Fig_Energyargument}(a) shows schematically the magnetization energy
$E_{mag}=\frac{1}{2\mu_0} (B-\mu_0 H)^2$ for $n=0$ and the oscillating dHvA energy described by the LK-formula $E_{osc}=a \cos(2\pi F/B)$ to its simplest approximation with $a$ the oscillation amplitude and $F$ the dHvA frequency. The sum of both energies as function of $B$ is shown for three different magnetic fields in Fig.\ref{Fig_Energyargument}(b)-(d). Usually, there is only one minimum in the total energy for a given applied magnetic field and the system will assume this value of $B$. However if the curvature of the $E_{mag}$ is smaller than the curvature of $E_{osc}$ two minima coexist at an applied field $H_1$ [Fig.\ref{Fig_Energyargument}(c)] and the induction will jump discontinuously from the value $B_1$ to $B_2$ when sweeping the magnetic field through $H_1$.

For a plate-like sample, oriented normal to the applied magnetic field
$\mathbf{H}$ ($n=1$), the boundary condition $B=\mu_0[H +
(1-n)M]=\mu_0 H$ is required even in the interval $B_1<\mu_0 H<B_2$.
Therefore, the induction $B$ can not change discontinuously and a
homogeneous state is no longer possible. The plate breaks up into
domains of opposite magnetization. The volume fractions of the
domains with the respective inductions $B_1$ and $B_2$ are adjusted
in a way that for the average induction of the sample
$\overline{B}=\mu_0 H$ is fulfilled~\cite{Condon1966}. The regions
with $B_1<\mu_0 H$ are diamagnetic, those with $B_2>\mu_0 H$ are
paramagnetic. The domain walls between the phases $B_1$ and $B_2$
run parallel to $\mathbf{H}$ across the plate. In contrast to
magnetic domains in common ferro- and antiferromagnetism, Condon
domains do not have their origin in the interaction of electrons via
their spin moments, but via their orbital motion.

For a sample shape with intermediate demagnetization factor,
$0<n<1$, the above mentioned interval of magnetic field $B_1<\mu_0
H<B_2$ with the occurrence of the instability will be reduced
compared to the plate-like sample with the field range of this
interval proportional to $n$. Therefore, samples of arbitrary shape
will still show the nonuniform domain state with the same dia- and
paramagnetic phases, $B_1$ and $B_2$, whose domain structure might
however be more complex.

Besides the analogy with the van der Waals gas, there is a close
analogy of the CDS with the intermediate state of type-I
superconductors, where the same boundary condition of a magnetic
field applied to a sample of nonzero demagnetization factor leads to
the formation of alternating domains in the normal and
superconducting state. Condon domains, however, have the unique
feature that the transition between the uniform and the
inhomogeneous domain state occurs periodically in subsequent dHvA
oscillations.

Equation~\ref{equation1} defines the boundary between the uniform
and the Condon domain state. The resulting CDS phase diagram in the
$(H,T)$ plane can be predicted by means of the Lifshitz-Kosevich
(LK) formula for the oscillatory magnetization of the dHvA signal
using the Fermi surface parameters, like the curvature
$A''=\partial^2 A/\partial k^2$ of the Fermi surface cross section
$A$ and the effective mass $m^{*}$, and the Dingle temperature
$T_\mathrm{D}$ as a parameter for the impurity-scattering damping of
the signal~\cite{Gordon2003}.

Up to now, Condon domains have been observed by different
experimental methods: by NMR~\cite{Condon1968}, muon spin rotation
($\mu$SR) spectroscopy~\cite{Solt1996,Solt2002} and, more recently,
by local Hall probes~\cite{Kramer2005a}. All experimental
observations have in common that two distinct inductions $B_1$ and
$B_2$ or an induction splitting $\delta B=B_2-B_1$ are measured at a
given applied field $H$ and temperature $T$. However, these
measurements yielded only a few points well inside the ($H,T$)
diagram where Condon domains exist that could be compared with the
theoretically predicted diagram. As a consequence, for example, the
data on beryllium obtained by $\mu$SR required new phase diagram
calculations with a modified LK-formula for the
susceptibility~\cite{Solt2001}. The exact determination of the CDS
phase boundary, where $\delta B$ approaches zero, is difficult and
time-consuming with a difference measurement of $B_1$ and
$B_2$~\cite{Solt1999}.

It was shown recently that a small hysteresis occurs in the measured
dHvA signal upon passing the CDS phase boundary~\cite{Kramer2005b}.
Due to the irreversible magnetization, an extremely nonlinear
response to a small modulation field arises in standard ac
susceptibility measurements. The out-of-phase signal and the third
harmonic of the pickup voltage rise steeply at the transition point
to the CDS. The threshold character of these quantities offers
therefore a possibility to measure a Condon domain phase diagram.
One should note that the third harmonic of the susceptibility is
commonly used as a very sensitive tool to detect phase boundaries
also of other systems like e.g. the vortex-glass transition in
superconductors~\cite{Klein1998}.

In this article we determine the Condon domain phase diagram for
silver using the third harmonic of the ac susceptibility for the
detection of the nonlinear magnetic response. It was shown
earlier~\cite{Kramer2005} that detailed calculations of the
magnetoquantum oscillations in silver based on the LK-formula are in
good agreement with experimental dHvA data up to 10~T. This is
certainly due to the nearly spherical Fermi surface of silver.
Expecting a good agreement with the theoretically determined CDS
phase diagram, we applied to silver this first detailed
determination of the phase diagram.

\section{Experimental}

The measurements were performed on a high quality silver single
crystal of $4.1\times 2.1\times 1.0$~mm$^{3}$. The sample was cut
from the same piece than the sample used for the direct observation
of Condon domains using local Hall probe
detection~\cite{Kramer2005a}. The sample preparation is described in
detail elsewhere~\cite{Gasparov1975,Gasparov1993}. The sample has a
residual resistance ratio
$R_{300~\mathrm{K}}/R_{4.2~\mathrm{K}}=1.6\times$10$^{4}$, measured
by the contactless Zernov-Sharvin method~\cite{Zernov1959}. The high
quality of the sample results in a very low Dingle temperature,
which was estimated from standard dHvA analysis to be about
$T_\mathrm{D}=0.2$~K yielding an electronic mean free path of about
0.8~mm.

A standard ac modulation method with a compensated pickup coil
system was used. Both pickup coils are identical and consist of
about 400 turns. A long coil wound by a copper wire produced the
modulation field with variable amplitude at frequencies of
$20-200$~Hz. The pickup voltage was simultaneously measured by two
lock-in amplifiers on the first and on higher harmonics. The
measurements were performed in a superconducting coil up to 16~T as
well as in a resistive coil up to 28~T at temperatures of
$1.3-4.2$~K. The long side of the sample was parallel to the
$[100]$-axis of the single crystal and was slightly tilted ($\sim
5$~deg) with respect to the direction of the applied magnetic field
so that only the dHvA frequency from the "belly" orbit of 47300~T
existed in the frequency spectrum.

The method of nonlinear detection, we use here, is applied to
determine for the first time a CDS phase diagram over a broad range
of temperatures and magnetic fields. Therefore, we will present
carefully the technical details of the measurements in order to show
the robustness of the phase diagram determination with respect to
changing experimental parameters and measurement conditions.

\section{Evidence of hysteresis in silver}
The employed method to determine a Condon domain phase diagram is
based on the appearance of hysteresis in the CDS which was first
discovered on beryllium~\cite{Kramer2005b}. Hysteresis appears at
the phase transition to the CDS and this results in some radical
changes in the response to an ac modulation field. In the following,
we will show that the characteristic nonlinear features in the ac
response, as found in beryllium, are also observed in silver.

\begin{figure}[tb]
\begin{center}
       \includegraphics*[width=0.75\linewidth, bb=138 518 468 792]{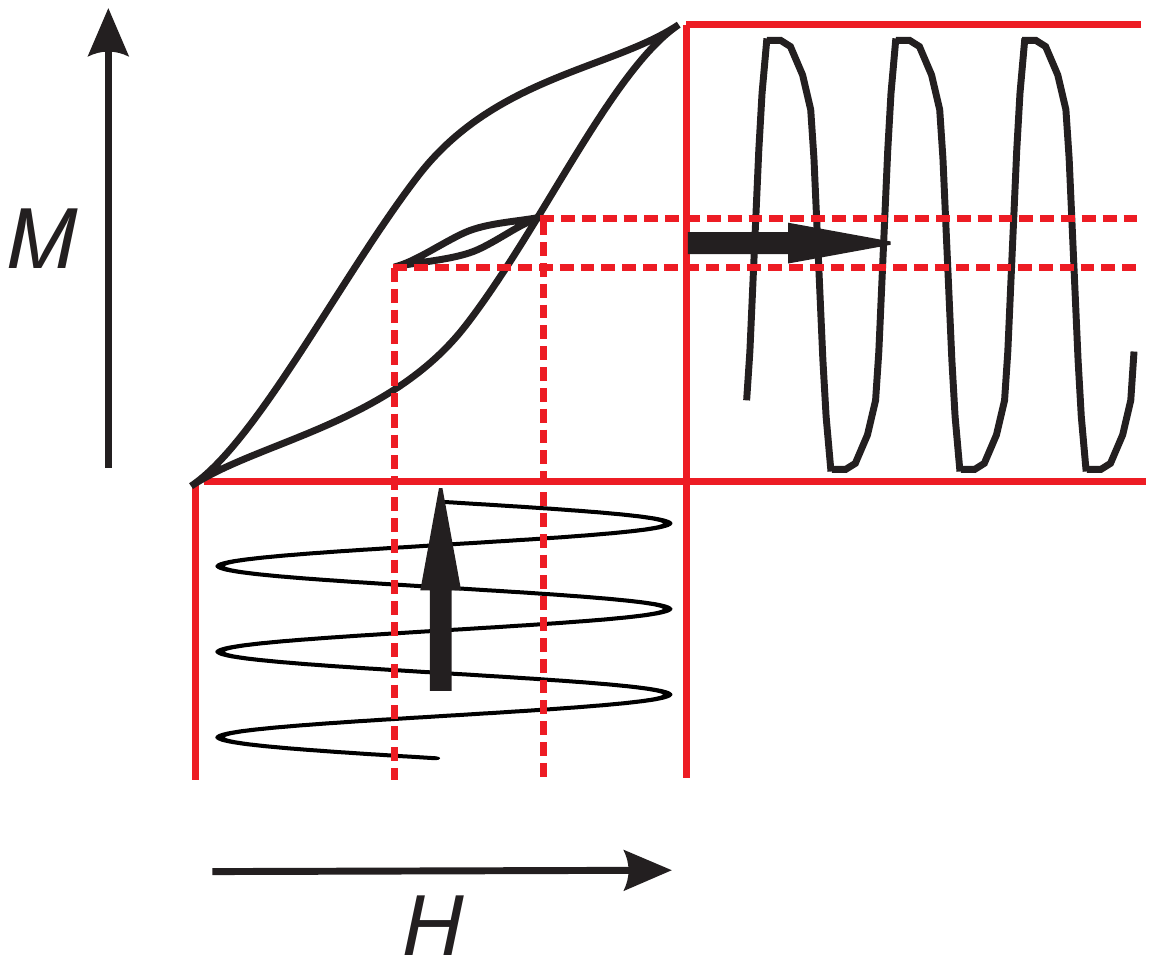}
       \caption{(Color online) Schematic representation of two minor hysteresis loops. The ac response
       upon field modulation is
       window-like and slightly shifted in phase.
       The normalized response (apparent susceptibility) decreases
sharply when the modulation amplitude becomes of the order
       of the hysteresis loop width.
       \label{Fig_hyst_3rd_schematic}}
        \end{center}
\end{figure}

In presence of hysteresis the amplitude of the susceptibility,
normalized on the modulation level, depends on the modulation
amplitude. This is expected to occur when the modulation level is
of the order of the hysteresis loop width. The schematic
representation of a hysteresis loop in
Fig.~\ref{Fig_hyst_3rd_schematic} explains this nonlinear response
to an ac field modulation.
As a result, after the transition to the CDS, the positive
(paramagnetic) part of the susceptibility turns out to be reduced.
From a comparison of two normalized susceptibilities, one measured
with high and the other with low modulation level, we can in
principle find where the amplitude reduction starts and thereby the
transition point to the CDS.

\begin{figure}[tb]
\begin{center}
       \includegraphics*[width=\linewidth, bb=5 7 214 302]{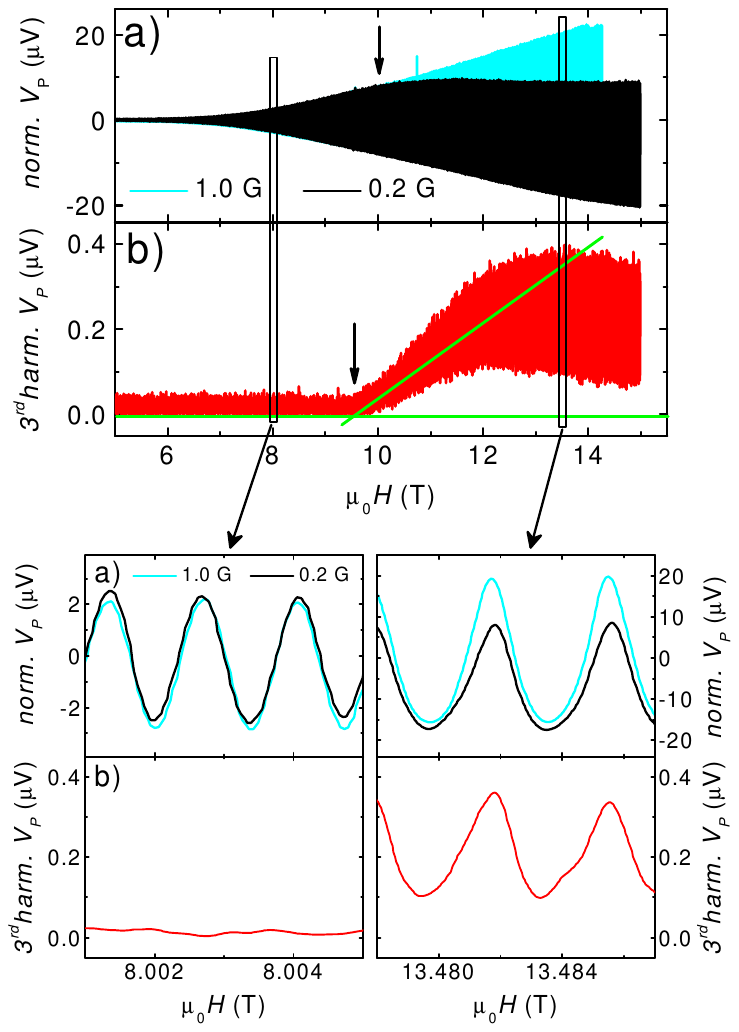}
       \caption{(Color online) (a)~Pickup voltage normalized on the modulation level
       for low and high modulation level. Up to about 10~T the
       response is linear with respect to the modulation level.
       (b)~Third harmonic of the pickup voltage measured at 0.2~G modulation
       amplitude showing that starting from 9.5~T the harmonic content in the
       response increases steeply. Lower part of the figure shows respective
       zooms. Both data measured at 2.7~K.
       \label{Layout1}}
        \end{center}
\end{figure}

Figure~\ref{Fig_hyst_3rd_schematic} shows as well that the response
to a sinusoidal field modulation becomes window shaped and is
slightly shifted in phase with respect to the input. Therefore, both
the third harmonic and the out-of-phase signal of the pickup voltage
increase steeply when the CDS phase boundary is
crossed~\cite{Kramer2005b}. This threshold behavior offers a simple
way to determine the transition point of the CDS. The major
advantage of third harmonic and out-of-phase part measurements is
that only one magnetic field or temperature sweep through the
transition is needed.

\begin{figure}[tb]
\begin{center}
       \includegraphics*[width=\linewidth, bb=5 7 218 220]{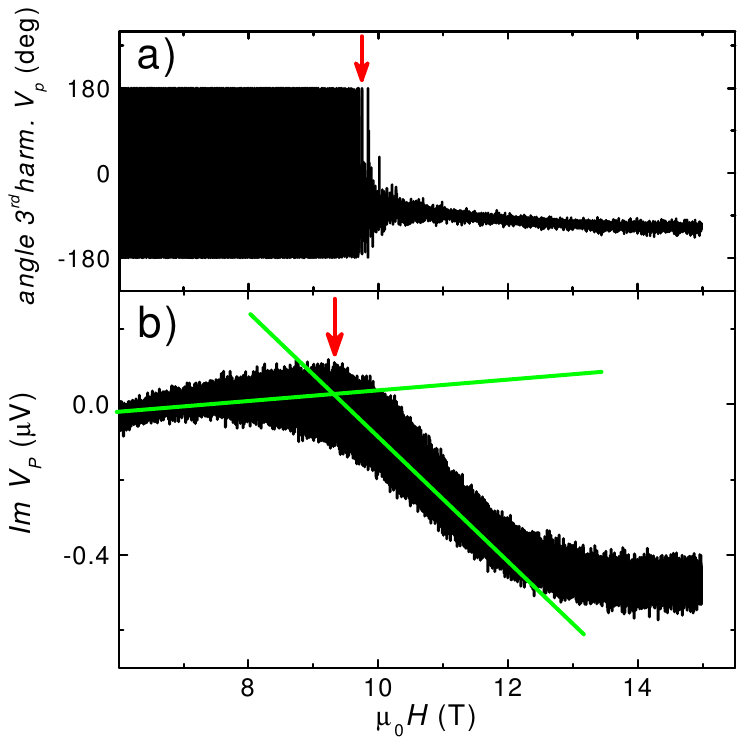}
       \caption{(Color online) (a)~Phase angle of the third harmonic showing clearly the
       transition between noise outside the CDS and a fixed phase in
       the CDS. (b)~The out-of-phase part of the first harmonic
       response changes due to the arising hysteresis.
       Both data measured at 2.7~K and 0.2~G modulation level.
       \label{Fig_phase_Im}}
       \end{center}
\end{figure}

Figures~\ref{Layout1} and \ref{Fig_phase_Im} show the above
discussed nonlinear features in the pickup signal at constant
temperature $T=2.7$~K measured in the superconducting coil.
Figure~\ref{Layout1}(a) shows two traces of the normalized pickup
voltage, i.e. the susceptibility, obtained in the same conditions
with 1.0~G and 0.2~G modulation amplitude at 160~Hz modulation
frequency. In principle, both modulation levels are small enough
compared to the dHvA period of about 20~G at 10~T that identical
traces are expected for the susceptibility. The expanded view around
8~T, which is outside the CDS, shows that the normalized signals are
indeed identical. For higher fields, on the other hand, the upper
part of the susceptibility waveform, measured with the smaller
modulation level, is reduced. The expanded view around 13.5~T shows
that the signals are identical except for the positive part of the
oscillation. This implies that at this part of the dHvA oscillation
the magnetization is irreversible and there is a small hysteresis
loop. The width of the hysteresis loop is of the order of 0.2~G.

A similar decrease of the normalized response was observed
earlier~\cite{Smith2004} on silver at low temperatures.
Unfortunately, because of the absence of the experimental
parameters, this study can be compared only qualitatively with our
data.

The magnetic field where the normalized pickup voltages start to
differ between low and high modulation level is marked approximately
by an arrow in Fig.~\ref{Layout1}(a). We obtain for the critical
magnetic field $\mu_0H_{c1}=10.0$~T.

Figure~\ref{Layout1}(b) shows the behavior of the third harmonic
which was simultaneously measured with the first harmonic response
in Fig.~\ref{Layout1}(a) for 0.2~G modulation amplitude. For
magnetic fields lower than the critical field there is only noise.
At the transition to the CDS hysteresis arises and the third
harmonic increases very steeply. This is nicely seen in the
respective expanded views. The critical field of the CDS phase
boundary can be obtained as the intersection of the two straight
lines shown in Fig.~\ref{Layout1}(b). Here, the critical field is
found as $\mu_0H_{c2}=9.6$~T.

The amplitude of the third harmonic is expected to go to zero in
each diamagnetic part of the dHvA period because the sample
magnetization is here homogeneous and without hysteresis. The
presented behavior in Fig.~\ref{Layout1}(b) does not go to zero
exactly which is certainly the result of a small rectification
effect or, what is the same, the result of phase smearing of the
oscillation signal. The homogeneity of the coil is about 30~ppm in a
sphere with 1~cm diameter which may result in a field inhomogeneity
of about 1~G in the sample volume at 10~T. Therefore, the transition
to the CDS does not occur simultaneously in the whole sample. This
effect will be much bigger in a resistive coil where the homogeneity
is 20 times worse. However, we will see below, that the third
harmonic rectification does not affect the determination of the
critical field of the CDS.

In Fig.~\ref{Fig_phase_Im}(a) the phase angle of the third harmonic
is shown for the same conditions like in Fig.~\ref{Layout1}. For
magnetic fields where the amplitude of the third harmonic is below
the noise level its phase angle is not determined. Therefore, the
phase varies between $-180$ to $+180$ degree. With the appearance of
a third harmonic signal at the transition to the CDS the phase
becomes finite. This passage has a threshold character as well. The
arrow in Fig.~\ref{Fig_phase_Im}(a) shows the position of the
threshold which yields the critical field $\mu_0H_{c3}=9.7$~T.

The behavior of the out-of-phase part of the first harmonic
response, shown in Fig.~\ref{Fig_phase_Im}(b), offers another
possibility to determine the critical field. In the uniform state,
without domains, the imaginary part is small and varies smoothly
especially at low magnetic field due to the magnetoresistance and
changing eddy currents. After the transition to the CDS the
out-of-phase signal changes rapidly. The transition point can be
found as the intersection of two lines, as it is shown in
Fig.~\ref{Fig_phase_Im}(b). Here, we obtain for the critical field
$\mu_0H_{c4}=9.4$~T.

A comparison of the values $H_{c1...c4}$ for the transition field
at 2.7~K shows that they are very close. We note that the critical
field of about 10~T agrees roughly with the phase boundary found
in the Hall probe experiments~\cite{Kramer2005a}. All above
presented methods could, in principle, be used to determine the
phase boundary of the CDS. The first method
(fig.~\ref{Layout1}(a)) requires at least two field sweeps.
Measurements of the out-of-phase part (fig.~\ref{Fig_phase_Im}(b))
are not precise, due to the high conductivity of silver and the
resulting eddy currents (The situation might be different in a
less conducting metal). Therefore, for silver the third harmonic
measurements to determine the CDS phase diagram are preferred.
Moreover, we will see below that the obtained values with the
third harmonic for the phase boundary ($H,T$) do not depend
drastically on the frequency and amplitude of the field modulation
which offers the possibility to measure in noisier conditions. The
found scattering in the values of the transition fields obtained
from the different methods gives an uncertainty of about $\pm
0.5$~T in the transition fields.

\section{Phase diagram}
Because of the increased noise level in the water-cooled resistive
magnets, we needed to increase the signal-to-noise ratio by using
rather high modulation frequencies $\approx$160~Hz and higher
modulation amplitudes, 1.0~G and more. In the following we check
whether the modulation frequency and amplitude can be varied without
changing the value of the critical field deduced from the third
harmonic response.

Modulation amplitudes of the order of the width of the hysteresis
loop are required to resolve the amplitude reduction in the
normalized pickup voltage in Fig.~\ref{Layout1}(a). For the third
harmonic signal, as shown in Fig.~\ref{Fig_hyst_3rd_schematic}, the
nonlinear features persist up to high modulation amplitudes.
\begin{figure}[tb]
\begin{center}
       \includegraphics*[width=\linewidth, bb=5 8 294 225]{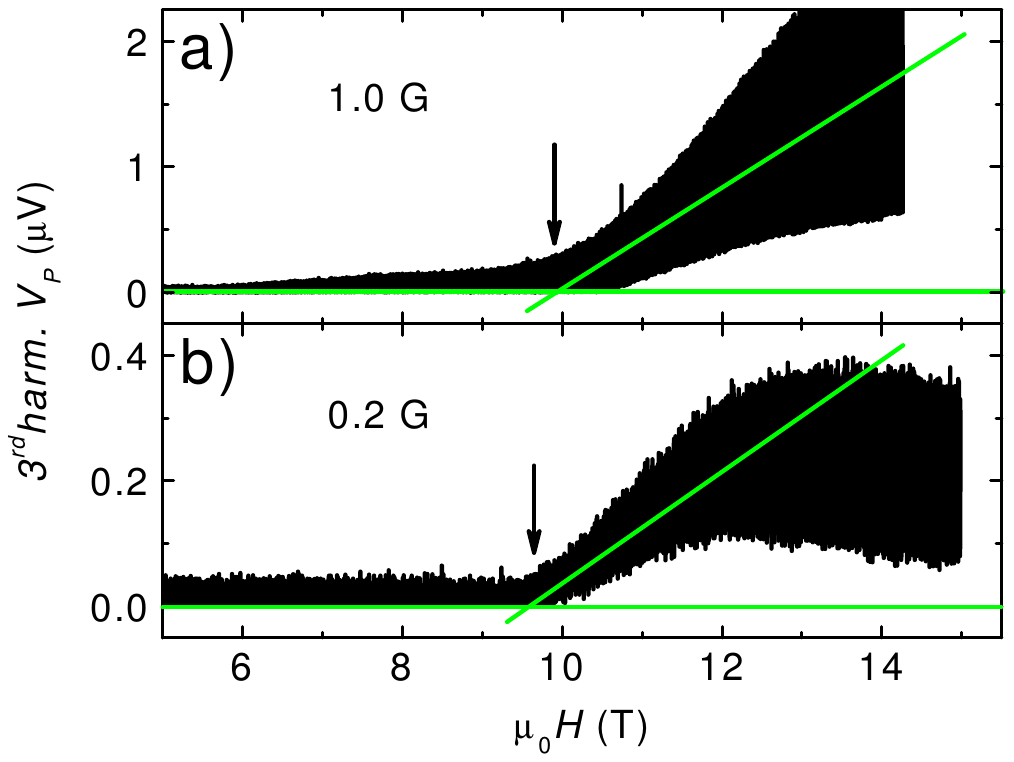}
       \caption{(Color online) Third harmonic response for two modulation levels
       measured in a superconducting coil at a modulation frequency of 160~Hz at 2.7~K.
       The same critical field is found from the steep increase.
       \label{Fig_3rd_amp_dep}}
        \end{center}
\end{figure}
Figure~\ref{Fig_3rd_amp_dep} shows traces with 0.2 and 1.0~G
modulation amplitude. For 1.0~G modulation amplitude there is a very
small third harmonic signal before the transition to the CDS takes
place. This small contribution to the third harmonic is due to the
nonlinearity of the dHvA effect itself~\cite{Shoenberg1984}.
Nevertheless, the position of the sharp increase remains unchanged.

\begin{figure}[tb]
\begin{center}
       \includegraphics*[width=\linewidth, bb=5 8 302 204]{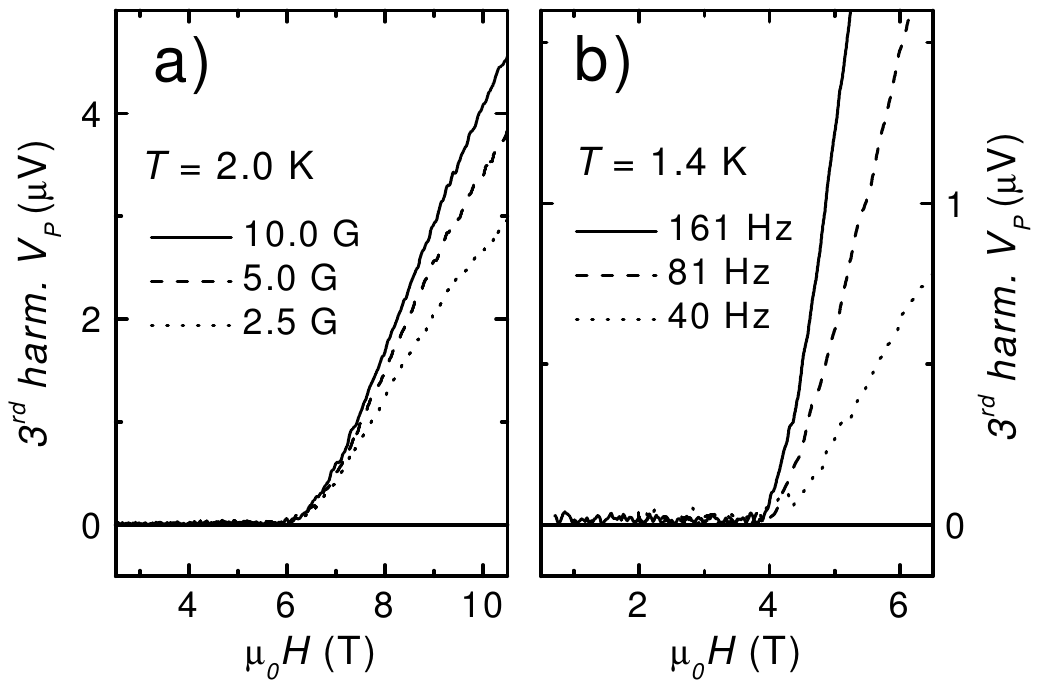}
       \caption{Third harmonic response measured after averaging over the
       oscillations in a resistive coil for different modulation
       amplitudes~(a) and modulation frequencies~(b).The critical field
       values obtained at a given temperature are independent on modulation frequency and
       amplitude.
       \label{Fig_resistive_freq_amp_dep}}
\end{center}
\end{figure}
Figure~\ref{Fig_resistive_freq_amp_dep} shows that increasing the
modulation amplitude up to 10~G and varying the modulation
frequency by a factor four between 40~Hz and 160~Hz does not
change the position of the critical field, either. The results
presented here were obtained in the resistive magnet. The
measurements were made at low temperatures in order to compare
them with data obtained in the superconducting magnet.

\begin{figure}[tb]
\begin{center}
       \includegraphics*[width=\linewidth, bb=0 3 264 202]{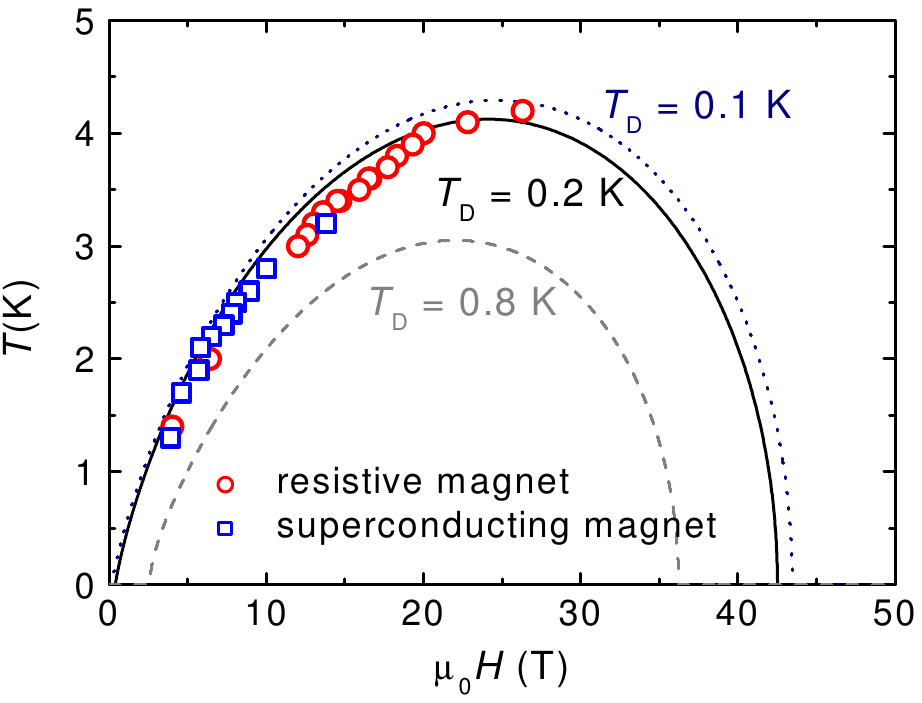}
       \caption{(Color online) Phase diagram in the $(H,T)$ plane for silver. Experimental
       points from the superconducting and resistive magnet. The
       solid line is the CDS boundary calculated by the LK-formula
       for $T_\mathrm{D}=0.2$~K~\cite{Gordon1999,Kramer2005}. The dashed and the dotted
       lines correspond to $T_\mathrm{D}=0.8$~K and $T_\mathrm{D}=0.1$~K, respectively.
       \label{Fig_phase_diagram}}
       \end{center}
\end{figure}

All results for the CDS transition points obtained in the
superconducting and the resistive magnets are presented in
Fig.~\ref{Fig_phase_diagram}. The critical fields for each
temperature are found as the field where the third harmonic response
starts to arise like in Fig.~\ref{Fig_resistive_freq_amp_dep}. One
should note that near the flat maximum of the phase diagram
$T$-sweep measurements would be in principle better. The solid line
in Fig.~\ref{Fig_phase_diagram} is the CDS boundary calculated for
silver using the LK-formula for the susceptibility criterion
$\chi=1$ with a Dingle temperature of $T_\mathrm{D}=0.2$~K for our
sample~\cite{Gordon1999,Kramer2005}. For comparison, theoretical
diagrams for $T_\mathrm{D}=0.1$~K and 0.8~K are shown in
Fig.~\ref{Fig_phase_diagram}.

A good agreement of the phase diagram predictions based on the
LK-formula with our data can be seen for a Dingle temperature
$T_\mathrm{D}=0.2$~K. Data points obtained in superconducting and
resistive magnets overlap which supports that the different
measurement conditions did not affect the precise determination of
the phase boundary.

\section{Conclusion}

Like previously observed in beryllium~\cite{Kramer2005b}, we have
shown that hysteresis appears in silver in the Condon domain state.
This substantiates that hysteresis is likely to occur in all pure
metals that exhibit Condon domains. The hysteresis leads to with
threshold character arising extremely nonlinear response to a
modulation field. In particular, the third harmonic response to a
modulation field increases sharply upon entering into the Condon
domain state. This offered the possibility to determine easily the
CDS phase boundary with high accuracy.

The critical fields obtained from the third harmonic of the pickup
signal of the ac modulation technique, turned out to be independent
on changes of the modulation frequency and amplitude. Due to this
independence this method could be used with higher modulation
frequencies in pulse magnetic fields.

Very good agreement of the CDS phase diagram is found with
calculations of the dHvA signal based on the LK-theory. This
agreement shows that the LK-formula describes well the field
dependent magnetization of the nearly spherical Fermi surface of
silver. Furthermore, the agreement demonstrates that the described
method is correct for the determination of the CDS phase diagram.

\begin{acknowledgments}
We are grateful to I.~Sheikin and V.~P.~Mineev for fruitful
discussions.
\end{acknowledgments}

\bibliography{REF_CondonDomain}
\end{document}